# An Empirical Study on Developing Secure Mobile Health Apps: The Developers' Perspective

Bakheet Aljedaani
*The University of Adelaide*
Adelaide, SA 5005, Australia
bakheet.aljedaani@adelaide.edu.au

Aakash Ahmad
*University of Ha'il*
Ha'il, Saudi Arabia
a.abbasi@uoh.edu.sa

Mansooreh Zahedi and M. Ali Babar
*The University of Adelaide*
Adelaide, SA 5005, Australia
[mansooreh.zahedi | ali.babar]@adelaide.edu.au

*Abstract* — Mobile apps exploit embedded sensors and wireless connectivity of a device to empower users with portable computations, context-aware communication, and enhanced interaction. Specifically, mobile health apps (mHealth apps for short) are becoming integral part of mobile and pervasive computing to improve the availability and quality of healthcare services. Despite the offered benefits, mHealth apps face a critical challenge, i.e., security of health critical data that is produced and consumed by the app. Several studies have revealed that security specific issues of mHealth apps have not been adequately addressed. The objectives of this study are to empirically (a) investigate the challenges that hinder development of secure mHealth apps, (b) identify practices to develop secure apps, and (c) explore motivating factors that influence secure development. We conducted this study by collecting responses of 97 developers from 25 countries – across 06 continents – working in diverse teams and roles to develop mHealth apps for Android, iOS, and Windows platform. Qualitative analysis of the survey data is based on (i) 8 critical challenges, (ii) taxonomy of best practices to ensure security, and (iii) 6 motivating factors that impact secure mHealth apps. This research provides empirical evidence as practitioners' view and guidelines to develop emerging and next generation of secure mHealth apps.

*Keywords—Empirical Software Engineering, Secure Software Development, Mobile Health, Software Engineering for Mobile*

## I. INTRODUCTION

Mobile health apps empower healthcare stakeholders (i.e., medics, patients, etc.) to exploit embedded sensors of a device for health diagnostics such as monitoring body temperature, pulse rate, and blood pressure [1]. mHealth apps rely on portability and context-sensitivity of mobile computing to improve access to healthcare services that are cost-effective, scalable, and pervasive [2]. World Health Organization (WHO) has defined mHealth as *"medical and public health practice supported by mobile devices, such as mobile phones, patient monitoring devices, personal digital assistants (PDAs), and other wireless devices [3]"*. mHealth apps range from decision support for reproductive health to fitness monitoring apps for nutrition management [4]. The usage of mHealth apps by healthcare practitioners and patients is on the rise with 350000 such apps available in two app repositories provided by Android and iOS platforms [5].

Mobile apps in general and mHealth apps in particular collect, process, store, and transmit user and device data in and out of a device over various networks [4]. Typical examples of such apps are location services or a banking app that needs device's type and ID, user's location and preferences to function properly [6-8]. Any compromise on the confidentiality, integrity, and availability of such data leads to severe consequences including but not limited to compromised devices, location and activity tracing, and financial loss. Therefore, security of mobile apps is a critical concern for users (data security) and app providers (secure app development) [9]. In the case of mHealth, security of mobile apps becomes a significant concern due to privacy and integrity of health critical data [4, 10] as an attack can modify the blood pressure or pulse rate of a patient that has medical, legal, and social consequences [11].

Existing research such as [1, 6, 11-14] indicate that security of mHealth apps lags behind the capabilities of adversaries and the sophistication of cyber-attacks that target the apps. A recent study suggests that despite their benefits, a wide spread adoption of mHealth apps is hindered due to users' concern about security and privacy of their personal and health critical information [15]. To address such issues, it becomes vital to investigate the security of mHealth apps by incorporating practitioners' view on challenges, best practices, and motivations to ensure secure mHealth apps. Such an investigation can help us to understand and address some fundamental issues such as why mHealth apps are not secure? and what efforts could be made to make them secure? In this paper, we investigate the security of mHealth apps based on developers'[1] view to understand (i) critical challenges that hinder development of secure apps (ii) development practices to ensure security of the apps, and (iii) motivations for secure app development. We outline the Research Questions (RQs) as below:

*RQ1:* *What challenges do developers face in developing secure mHealth apps?*

*RQ2:* *What practices are used to incorporate security measures in mHealth apps?*

*RQ3:* *What motivates developers to engineer and develop secure mHealth apps?*

We conducted a web-based survey that received responses from 97 app developers. Demography analysis of developers' data indicates their experiences, team size/dynamics, and professional roles to develop mHealth apps for mobile platforms including Android, iOS, and Windows. To ensure the quality of responses, we only allowed developers with first-hand experience of developing mHealth apps to participate in our survey. The results can be beneficial for researchers and practitioners (e.g., mHealth app developers, managers, research engineers) to support research and development of emerging and next generation of secure mHealth apps.

## II. RELATED WORK

We organized the related work in three themes (i) security issues, (ii) development challenges, and (ii) developers' practices and motivations for secure app development.

---

[1]The terms *developer*, *practitioner*, *respondent*, and *participant* have been used interchangeably in this paper all referring to professionals who are engaged with engineering and development of secure mHealth mobile apps.

## A. Security Issues with Current mHealth Apps

A recent mapping study [16] of 365 papers on m/uhealth security and privacy highlights that the security and privacy specific education and training for developers is one of the critical factors to support secure development of mHealth systems. Some other studies have focused on investigating security of mHealth apps by means of assessment experiment [1], static and dynamic analysis [6], comparative analysis [11], security assessment [12], and vulnerability scanning [13]. These studies revealed that most of the mHealth apps lack the incorporation of security countermeasures, such as access control and threat detection. For example, He et al. [14] revealed that several mHealth apps are prone to attacks due to inherent vulnerabilities such as transmitting mHealth data in unencrypted form and logging sensitive information. Moreover, numerous mHealth apps were found to have component exposure threats and some apps store unencrypted information on external storage, e.g., SD Card, where a malicious app can read and modify it [14]. In [6], the authors have reported that mHealth apps developers may fail to appropriately implement even the basic security solutions such as authentication, access control, and data encryption that impacts security and privacy of health information.

## B. Challenges to Develop Secure mHealth Apps

Through a systematic review [17], we identified that in order to ensure security of mHealth apps throughout their development lifecycle, the developers need to be vigilant about security-related issues. Contrary to the finding of [16, 17], some studies highlight that mHealth apps developers are not even familiar with fundamental solutions such as security measures [13, 18], security tools [8], and trusted libraries [12, 19] that are critical for secure apps development. Developing mHealth apps needs updated Security Knowledge (SK), especially knowledge about connecting mHealth apps technologies that are driven by context-sensitive and pervasive computing enabled by mobile and Internet of Things (IoT) [14]. It is vital to mention that the primary purpose of health and security regulations (e.g., Health Insurance Portability and Accountability Act - HIPAA) is to protect health critical data and to promote trustworthy systems. However, such regulations do not provide explicit guidelines, processes, and methods to assist the stakeholders (e.g., users, developers) of mHealth apps [2, 10]. Moreover, frameworks and standards for developing secure mHealth apps are still missing from the existing regulations, policies, and development initiatives [6, 11, 13].

During software development, project specific constraints such as time and cost to deliver, can pressurize developers to focus on satisfying functional specifications first and patching advanced security issues after initial release [4, 18]. Patching security specifications in a released app is a costly approach and introduces new vulnerabilities after fixing the existing issues [1]. Also, the lack of security testing due to development project constraints for mHealth apps is a challenge that has been reported [1, 18, 20]. Other issues such as absence of security experts, lack of understanding for security testing tools, and shortage of budget to conduct app testing are attributed as fundamental challenges for developing secure apps.

## C. Developers' Practices and Motivations

A lack of adoption or poor understanding of security practices by mHealth apps developers' hinders the development process for secure software engineering [3]. The consequences of compromised security practices – during software development lifecycle (SDLC) – makes apps vulnerable, such as permitting an app to share health critical data with other mobile apps (e.g., untrusted apps or external untrusted hosts [2]). Thamilarasu et al. [18] examined top 15 Android-based mHealth apps and found 248 vulnerabilities. The study identified that top 3 vulnerabilities were caused by development practices followed by developers, e.g., selection of cipher method or implementation of specific algorithms to request or grant permissions. It concludes that most of these vulnerabilities could have been prevented by adopting development practices that adhere to secure SDLC.

Developers' and team motivation are two of the key success factors for software projects [21]. Many security incidents are primarily caused by human rather than system failures [22]. Motivational factors that help mobile apps developers to achieve security were discussed by Weir et al. in [23]. The study concluded that developers were motivated by their SK and experience, considering security as a task that needs to be done in the right way, impacts of developing insecure software, and pursuing secure development as an enthusiasm. Developers motivation for secure software development when coupled with security specific education and training is among the most important factors for security of mHealth apps [17]. Reusing source code from a previous project is a common practice to promote reusability and cost effectiveness; nevertheless, security bugs can be inherited [11-14]. Also, copying or reusing the source code from web-based public repositories (e.g., Stack Overflow - SO, GitHub) without further examination of the source code is a common practice among app developers that leads to developing vulnerable apps. The impact of copy/pasting of source code from SO to Android apps has been investigated in [24]. The study revealed that 97.9% of the copy/pasted code contained at least one insecure code snippet.

To the best of our knowledge, there is no empirical study to accumulate the evidence of developers' perspective to investigate the challenges, practices, and motivations for developing secure mHealth apps. Our study provides empirical basis for the development practices that enable or enhance secure app development for mHealth systems.

## III. RESEARCH DESIGN

In this section, we detail four phase research design to conduct and document the proposed study. Each of the research phases is detailed below and illustrated in Fig. 1.

### A. Phase I - Study Protocol and Questionaire

We designed an online survey based on the findings of our systematic review [17] and guidelines by Kitchenham and Pfleeger [25]. As in Fig. 1, we utilized an online platform (Google Forms) that is easy to share, view, and manage across platforms. In the survey preamble, we briefly described the purpose and eligibility of our study. The survey contained 14 Questions (**Q1** to **Q14**) designed to answer RQs (Section I). The link to online survey is provided in [26].

**Q1-Q3** were screening questions to ensure that we engage developers with experience of mHealth apps. Selecting the option of zero apps redirected the respondent to submission section, indicating that s/he was not eligible to participate in the study. Respondents were asked to provide the name and link for at-least one mHealth app to ensure their eligibility.

**Q4-Q9** were demography specific questions including but not limited to years of experience, team size, work pattern.

**Q10-Q12** aimed to understand the challenges for secure mHealth apps. In Q10, we proposed a few statements using Likert-scale questions to rate the respondents' agreement to the reported challenges [17]. Q12 aimed to find out the practices by respondents to ensure the security of their apps.

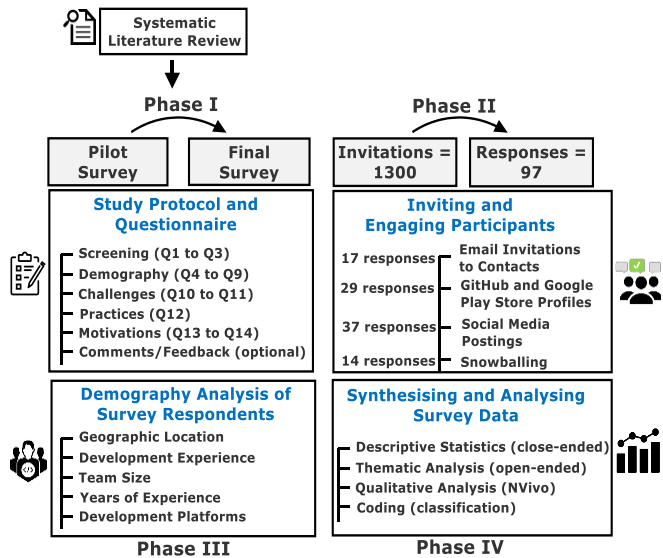

Fig.1. Overview of the Research Design

**Q13-Q14** aimed to identify the developers' motivations to develop secure mHealth apps. A few motivating factors were proposed in Q13 to rate respondents' agreement. We also provided an **optional question** at the end of our survey to allow respondents to share their comments/feedback.

**Pilot and Final Survey**: It should be noted that the survey questions were slightly modified (e.g., adding questions and removing ambiguity) during response collection. First, we revised the survey after 05 responses and second revision happened after19 responses. Our survey was designed in English and took around 15 minutes to complete. Our study is approved by the Human Research Ethics Committee at the *University of Adelaide (H-2019-115)*.

*B. Phase II - Inviting and Engaging the Participants*

Since this study sought specific type of developers (i.e., Q1-Q3, Screening in Phase I), we used a different method to ensure that all respondents meet our selection criteria. As in Fig. 1, first, we emailed our contacts having experience with developing mHealth apps and asking them to participate in the survey. Secondly, we extracted the email addresses of mobile apps developers from GitHub and Google Play which were publicly available in their profiles. We sent an invitation to over 1300 mobile apps developers. Thirdly, we posted our survey on several mobile apps developers' groups on prominent social media platforms (LinkedIn, Facebook, and Reddit, etc). Finally, we followed the snowballing method to reach more respondents. We obtained 97 valid responses. We obtained 37 responses from social media platforms, 29 from GitHub and Google Play users, 17 from personal contacts via email invitations, and 14 through snowballing technique. The details of the responses from the participants are available at [29].

*C. Phase III – Demography Analysis of Survey Respondents*

We now present some important demography analysis of the survey respondents that is also referred to as developers' information, as illustrated in Fig. 2. The demography analysis presented in Fig. 2 complements the survey responses (**Q1-Q14**) and helps us with fine-grained analysis of the survey results. For example, the platform(s) used for development in Fig. 2 (b) can help us understand the most and least preferred mobile computing platforms by developers to deploy their mHealth apps. Specific mobile platforms may pose different development challenges and encourage developers to adopt different practices for app development [1, 4]. In addition to the details in Fig. 2, 80% of the respondents had experience as full-time and 20% as part-time developers. All respondents were involved in mHealth apps development in various roles including but not limited to software engineers, project managers, technical leads, system architects, developers, and designers. The full details of the respondents' professional roles are available in [26].

*D. Phase IV – Synthesising and Analysisng Survey Data*

We analyzed the answers to the close-ended questions (e.g., Likert-scale) using descriptive statistics. For open-ended questions, we used thematic analysis method [27] to identify any recurring themes in the gathered responses. Thematic analysis supports extracting the data and synthesizing the results. We enhanced our analysis by using NVivo software, a popular computer-based tool, to organize and analyze data. The coding was initially done by one of the researchers in the team that was reviewed and revised (wherever required) by third researcher to avoid potential bias.

**Findings of the study**: We now present the findings of the survey to answer to outlined RQs. Answers to the RQs are presented in the dedicated sections as: (i) *security challenges* (**RQ1**) in Section IV, (ii) *development practices* (**RQ2**) in Section V, and (iii) *motivating factors* (**RQ3**) in Section VI.

IV. CHALLENGES OF SECURE MHEALTH APP DEVELOPMENT

In order to identify the challenges faced by developers, i.e., answer to RQ1, we formulated 8 statements, (**SC1** to **SC8**) based on the findings and guidelines of our SLR [17]. The statements capture the input of 97 respondents (**R1** to **R97**) as illustrated in Fig. 3. For an objective interpretation and assessment, we sought developers' feedback on each statement using five-scale Likert (*Strongly Agree*, *Agree*, *Not Sure*, *Disagree*, and *Strongly Disagree*) as in Fig. 3. Furthermore, the responses to these statements were complemented by an open-ended question that allowed developers to spontaneously share other challenges based on their knowledge and experience. The results of developers' perspective on each of the 8 challenges are detailed below.

**SC1) Insufficient SK of the developers:** Developing secure mHealth apps require proper SK. Typical examples of SK include but are not limited to coding practices, robust usage for security testing tools, and confidence in using third-party libraries that should be known by mHealth apps developers [16, 17]. We asked the respondents to rate their agreement or disagreement while considering insufficient SK as a challenge for developing secure mHealth apps. While 20% of the respondents disagreed, 80% of them affirmed that inadequate SK is a challenge for mHealth app development reflected as SC1 in Fig. 3. Some respondents elaborated that poor skills for

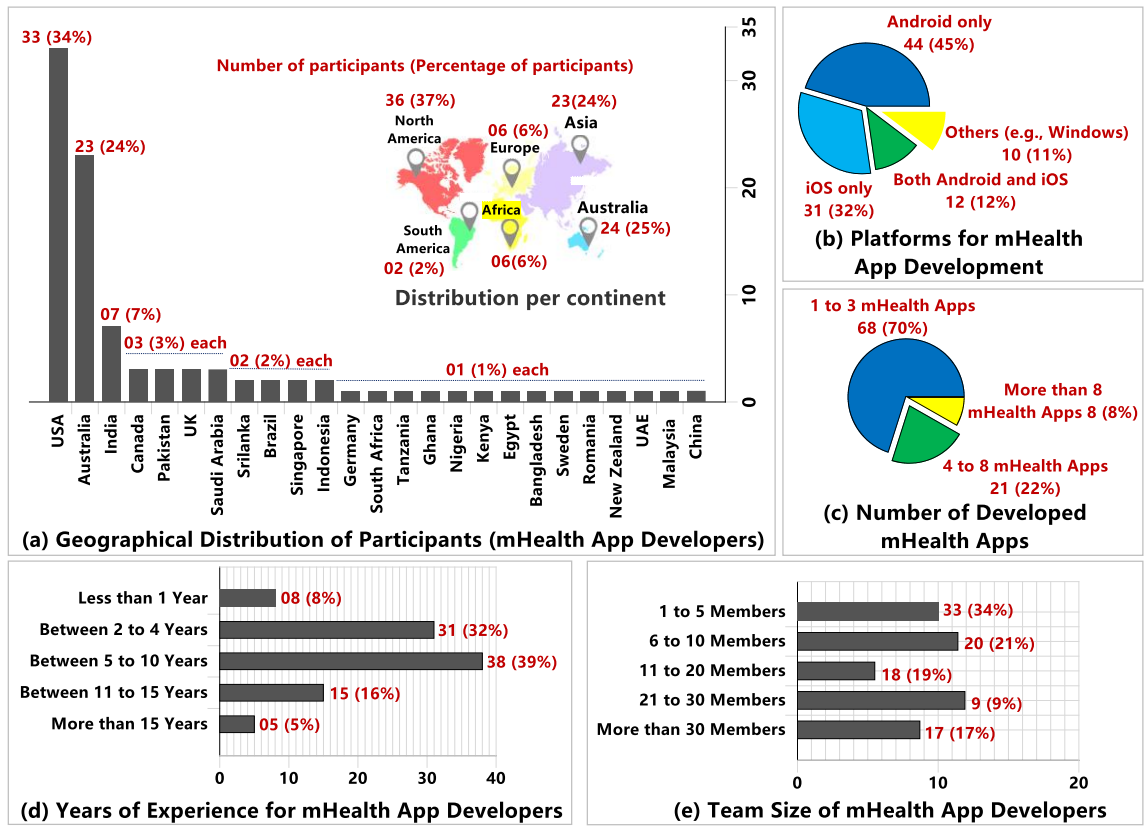

Fig. 2. Demography Analysis of the Survey Respondents (mHealth App Developers)
[*(a) Geographical Distribution, (b) Development Platforms, (c) Number of Developed Apps, (d) Years of Development Experience, (e) Team Size*] secure programming is a critical challenge to develop secure apps. It was highlighted that mHealth apps developers face difficulties in using programming tools and reusing code that can be vulnerable. For example, as per the claims of **R13** and **R45** *"Cloning of certain features by other developers"* and *"Picking up code written by the hacker represents a critical challenge"*. Some respondents pointed out that they have a lack of SK in dealing with attacks and vulnerabilities. Specifically, **R59** shared that *"[him/her] not being aware of potential security risks"* and **R46** emphasized that *"mHealth apps developers need to put more effort to enhance their SK through R&D and learning new technologies"*.

**SC2) Little or no budget for employing security:** Developing secure software requires allocating a specific budget to be spent on enhancing security of developed and deployed apps [17]. We asked our respondents to rate their agreement or disagreement with SC2, i.e., little or no budget for employing security. 85% of them affirmed that developing secure mHealth apps cannot be achieved without assigning an adequate budget. Only 15% disagreed with the statement SC2 in Fig. 3. Our analysis of the responses revealed that lack of allocated budget for secure mHealth apps represents the most critical challenge.

**SC3) Lack of involvement of security experts during software development:** The absence of security experts is being recognised as a factor that directly influences secure software development [28]. Involving security experts during apps development would give inexperienced developers access to the required SK (SC1) [17]. We asked our respondents if they consider lack of security experts' involvement during the development as a challenge as SC3 in Fig. 3. Given the responses, 70% of them lacked security experts within the team during mHealth app developmen; 30% disagreed with this statement. We looked into the experiences of the respondents, who showed their disagreement from Fig. 2 (d) and data in [26]. We found that the majority of the respondents (20/27 respondents, i.e., 74%) have more than 5 years of experience in developing mHealth apps. Hence, they depend on their ability and expertise in developing secure mHealth apps without relying on security experts.

**SC4) Poor security decisions during the development process:** Developing mHealth apps requires making appropriate security decision with respect to storing and using health critical data [22, 24]. Poor security decisions during mHealth apps development would leave mHealth apps vulnerable to security threats. Besides, with the high cost of fixing flaws (e.g., security patching [4]), mHealth apps development organization may face technical and legal issues for data breach. Thus, mHealth apps developers' and their organizations need to pay careful attention to security-related decisions. We asked respondents about their view on security decisions during the development process, SC4 in Fig. 3. 66% agreed that making poor security decisions during the SDLC is a challenge, while 30% disagreed with this statement. Only 4% of our respondents were neutral.

**SC5) Assumption about security issues resolved by app testers:** Ensuring security is a task that needs to be considered by software developers throughout the SDLC. However, Xie et al. in [29] concluded that one of the main reasons that leads software developers to make security errors is relying on others, such as app testers. In fact, 66% agreed that app testers should make decisions about ignoring or identifying security flaws. 33% disagreed and only 1% were neutral, as SC5 in Fig.3.

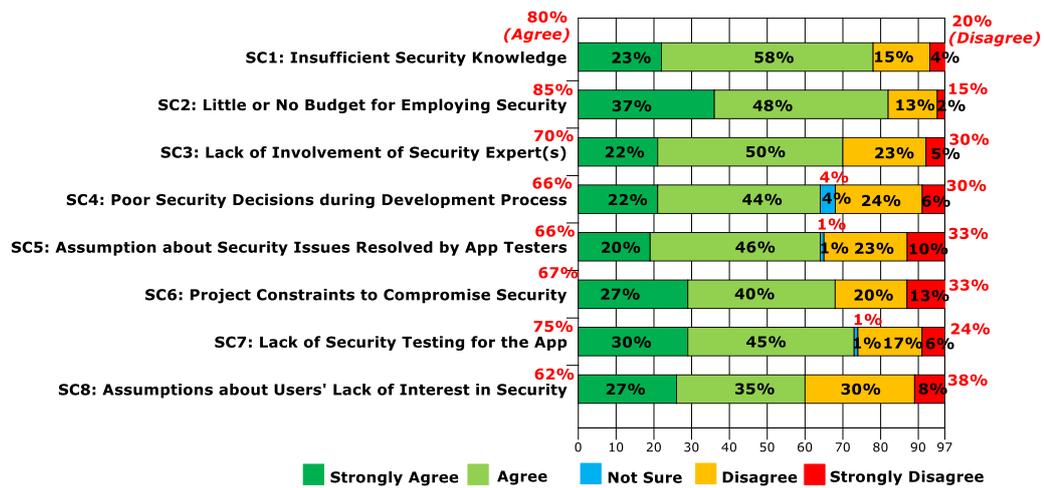

Fig. 3. Responses for Critical Challenges Related to Security of mHealth Apps.

**SC6) Project constraints to compromise security:** Project constraints, specifically time to deliver and cost efficiency (SC2), represent common challenges for software engineering projects as well as for SDLC of secure mHealth apps [28]. Time to deliver – relative to market competition – forces development teams and developers to primarily focus on functional requirements and often overlooking or ignoring quality requirements that include security features. According to our respondents, 67% of them agreed that meeting app development deadlines pressurizes developers to compromise security requirements and their resting. 33% disagreed with this statement SC6 in Fig. 3. Respondents including **R12**, **R37**, **R50**, **R58**, and **R62**, commented that it is a challenge to ensure the security of mHealth apps due to time constraints. **R62** climed that *"Lack of time for app delivery stresses developers that impacts quality specific requirments such a security, performance, scalability and many more. In secure [mHealth] app project deadlines is a sigificant challenge for developers"*.

**SC7) Lack of security testing:** Security specific testing is a crucial phase throughout SDLC to identify potential vulnerabilities in security and privacy of developed app. In the context of mHealth apps, it helps to address security threats, such as unauthorised access to health data, tampering with health data or reporting invalid data to health providers. We asked our respondents if they consider lack of security testing as a challenge to develop secure mHealth apps. While 75% of respondents agreed with this statement, 24% indicated their disagreement and 1% were neutral for SC7 in Fig. 3. Five respondents **R38**, **R45**, **R52**, **R62**, **R95** commented that the lack of rigorous testing for mHealth apps is the most critical challenge for them.

**SC8) Assuming that users are not that much interested in security**: Security should be incorporated and addressed ideally throughout the SDLC from requirement analysis to deployemt [30]. Incorporating security at later phases of software development or after release in the form of security patches can be costly exercise and can introduce new vulnerabilities [31]. However, the trade-offs between security and other quality attributes such as usability, perfromance can be problematic. **R27** and **R63** indicated that app accuracy and performance are challenges that affect the development of secure mHealth apps. Consequently, developers might assume that users are not concerned about security and hence lower the priority given to security due to budgeting or time constraints. 62% of the respondents agreed that users are not interested in security. 38% disagreed as per SC8 in Fig. 3. Four respondents **R54, R62**, **R73**, **R81** indicated that mHealth apps developers' pay little or no attention to security. **R62** stated that *"We are yet to be engaged in a mobile health project where we put specific focus on security aspects of the app"*. **R54** claimed that *"There [are] very few developers in my community who bother about security"*, and **R73** indicated that *"I think a large number of the other players in the space don't take it seriously enough. I think they aren't paranoid enough"*.

**Other Challenges (Open-ened):** In addition to close-ended statements from **SC1** to **SC8**, the respondents highlighted '*other*' challenges, detailed below, they see as important but were not presented as survey questions.

*a) Dealing with legal obligations, policies and procedures:* mHealth apps can be classified as low-risk apps (e.g., fitness apps) and high-risk apps (e.g., clinical decision-support apps) [32]. Given the criticality and confidentiality of health critical data, various governments have established different compliances such as HIPAA to ensure security and privacy. In this context, **R39** and **R57** highlighted that policies and regulations for mHealth app development are ideal but complex to implement and abide. Five respondents **R9**, **R55**, **R60**, **R72**, **R91** found lack of guidelines, documentation, and better procedures to ensure app security.

*b) Challenges of maintaining mHealth app and data:* Six respondents **R16**, **R29**, **R43**, **R57**, **R86**, **R92** indicated that there are difficulties in maintaining mHealth apps and data due to the complexity and privacy-preserving nature of data. Specifically, **R16** stated that maintaining mHealth apps requires proper security management team. **R57** suggested to keep third party libraries updated to avoid any security vulnerabilities from code exectuion. **R29** pointed out that

*"continuous updates [are needed] so that you don't miss any new medications or information"*. **R43** and **R92** mentioned that mobile phone and platform compatibility (e.g., Android, iOS etc. platforms) is a challenge in maintaining mHealth apps with **R82** indicating that *"Privacy preservation of the data collected by the app"* is the most critical challenge.

## V. PRACTICES TO ENSURE SECURITY OF MHEALTH APPS

In order to identify the developers' practices, i.e., finding answer to RQ2, we explored how security is integrated during mHealth apps development process by our respondents. We used open-ended questions because we did not want to limit our respondents while sharing the practices they adopt and their relative experiences [29]. Although there are several guidelines for Secure Development Lifecycle such as [33], we used Microsoft secure software development process (5 development tasks) to analyse and document the provided responses by the developers. Based on the developers' responses, we created a taxonomy of the development practices in Fig. 4 to classify the specific practices that ensure security during each task of a SDLC. The taxonomy in Fig. 4 also helps with conceptualisation and quick identification of all the practices that developer perceive as effective to enable or enhance app security.

**Task I – Requirements Engineering:** As the intial task of SDLC, requirements engineering involves identifying, specifying, managing, and implementing security requirements for the app to be developed. Engaging stakeholders in security requirement enginering is being recognised as a key to software success, as well as getting effective outcomes. In Section IV (SC4, Fig. 3), we reported that poor security decisions could be caused by lack of stakeholders' involvement in SDLC. Three respondents identified as **R60**, **R78**, **R96** indicated that they involve users feedback and try to negotiate requirements with users as an approach to employing security during requiremnets engineering. For example, the respondent **R78** claimed that *"In my case, it [requirements engineering] involved discussing requirements and what would be the most sensible solution to ensure security requirements"*.

**Task II – Software Design:** This task is essential to represent the identified security as a design or blueprint of the software to be implemented. **R45**, and **R60** indicated that selecting the right development platform and supporting tools would help to enhance the security of mHealth apps. Four respondents **R28**, **R60**, **R82**, **R85** mentioned that adhering to security guidelines helps ensure security of mHealth apps. **R45** and **R48** pointed out that as a design consideration, minimising data collection, sharing and asking for personal information positively impacts security of mHealth apps. **R51** and **R52** indicated that they ensure the security of mHealth apps through utilising a layered approach (incorporating a dedeictaed security layer) and secure data storage. Also, using security frameworks, standards along with policies would help to design secure mHealth apps and comply with security regulations. Furthermore, **R57** and **R72** indicated that they analyze and map the attack surface. **R72** claimed that *"Thinking about all ways where hackers could be intrusive can help design possible scenarios to countermeasure security breaches"*.

**Task III – Software Implementation:** This task of SDLC is about writing the source code to implement the design (Task II) based on the identified security requirements (Task I). App implementation focuses on coding/implementing security measures (e.g., encryption, anonymisation) into apps. **R57** and **R74** indicated using trusted libraries or APIs as a best practise to ensure the security of mHealth apps. **R57** suggested that he is *"Following industry specific [code libraries, APIs] practices rather writing custom cource code to ensure app security"*. **R74** claimd that *"[…], we [as part of development team] just do basic AES encryption of user data"*. **R30** and **R38** mentioned that they utilise static analysis [6] of source code as part of implementation task to ensure the security of mHealth apps.

**Task IV – Software Verification:** It aims to analyse the implemented app (Task III) and identify potential flaws that can be exploited for malicious access. As part of software verification, app testing is an effective approach to determine security vulnerabilities and threats, thus, employing suitable measures [4, 10]. Analyzing the responses, we found that 16 respondents have indicated app testing as their best practice to ensure security. Specifically, app testing involves practices such as attack simulation, data flow testing, code review, and penetration tests. **R86** mentioned that, *"Testing app vulnerabilities by simulating attacks is an effective practice to assess security strength of the application"*. **R84** and **R90** mentioned that they conduct data specific testing (e.g., data leakage testing, data encryption testing). **R83** and **R97** mentioned internal code review as a proven practise to test security strength of their apps. **R83** suggested about, *"Multiple people to review source code [can avoid potential bias of code inspection]"*. Five respondents **R9**, **R39**, **R57**, **R73**, **R89** indicated that they performed testing with external security experts. **R9** mentioned *"Hiring a computer security firm to run penetration tests"*, and **R89** suggested *"Invite third-party companies for security audit and testing"*.

*Task V – Software Deployment*: As the last task of SDLC, it refers to the release and deployment of the app to be used by users. One of the respondents, **R37**, indicated that they perform a final security review to ensure the security of mHealth apps. One respondent **R4** believed that most of the developers are only able to fix security flaws after deployment and user testing. **R4** claimed that, *"[in our team] we found that most of our peers approach security testing during deployment, once the system is operational"*. Furthermore, automatic vulnerability patching is an approach to ensure the security of mHealth apps. The overall responses regarding security practices suggested:

a) *Security-aware SDLC:* Considering security throughout a SDLC is a best practice to develop secure software. This practice not only ensures an incremental security (from requirements to deployment) but also reduces the cost and efforts of fixing security errors afterwards in the form of security patches. Five respondents **R26**, **R55**, **R73**, **R74**, **R85** indicated that throughout the SDLC satisfying security requirements of mHealth apps remain their primary concern. **R74** suggested that "[Development] should be more concerned about security before as cyber attacks frequency increases", and **R26** suggested, "Preserving user personal information from any leakage is ultimate success of any secure development process". Also, **R39** and **R57** indicated that they consult security experts throughout SDLC.

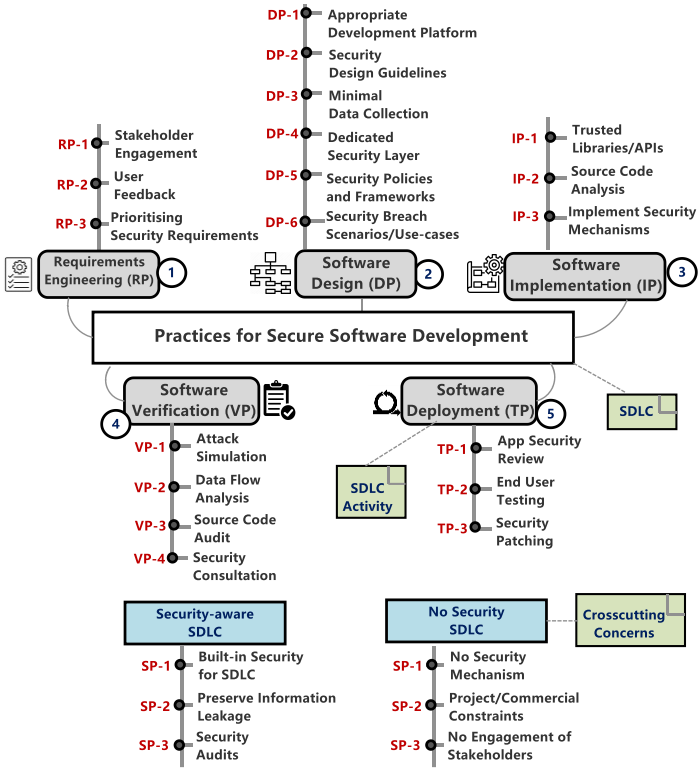

Fig. 4. Taxonomy of Development Practices for Secure mHealth Apps.

*a) No security approach is being considered:* While security becomes an important concern during SDLC, yet according to a few respondents, it still not much of a concern to them. Four respondents **R62**, **R70**, **R71**, **R81** indicated that they did not follow any approach to address security during the development process. **R61** claimed that *"We are yet to be engaged in a mobile health project where we paid specific focus to the security aspect",* and **R70** claimed that *"No process most of the time. Development teams [due to project constrainst and commercial reasons] have no direct link with medical teams (stakeholers), and due to this lack of interaction, user are using apps that can be disastrous for security and privacy of mHealth data".*

## VI. Motivations to Ensure Secure App Development

To identify the motivating factors for secure mHealth app development, i.e., to answer to RQ3, we formulated 6 statements (SM1 to SM6) based on the findings and guidelines of our SLR [17]. The statements capture the responses of 97 respondents (R1 to R97) as illustrated in Fig. 5. To objectively interpret and assess the responses, we sought developers' feedback on each statement using five-scale Likert (Strongly Agree, Agree, Not Sure, Disagree, and Strongly Disagree) as in Fig. 5. In addition, input to these statements was complemented by an open-ended question to allow developers to share other motivating factors based on their knowledge and experience.

**SM1) Security Leader in the Team to Influence the Development of Secure mHealth Apps:** Commitment to improving the security is one of the main goals that security team leads aim to achieve. The respondents **R2** and **R38** indicated that they follow security experts' comments and guidelines to ensure security. Specifically, **R2** claimed that *"Following security department feedback is required during the design"*, and **R38** sugested that *"[...] and following the guidelines of Security Expert"*. 59% of the respondents agreed that a security specialist as team leader influences and motivates them to develop secure mHealth apps. 36% disagreed and 5% remained neutral on the role of security leaders in the team as SM1 in Fig. 5.

**SM2) Secure Development to Maintain Vision and Reputation of the Organization:** Creating and maintaining the reputation is a common goal that software development organizations seek to fulfil their vision. **R45**, **R69**, **R91** emphasised that they consider maintaining organization reputation's, building trust and credibility as their motivations to develop secure mHealth apps. **R45** claimed that *"[...] and the second thing is by developing a secure application, the client's trust will become strong"*. Satisfying end users can also help to maintain the organization's reputation and fulfil its vision. Four respondents **R27**, **R61**, **R72**, **R92** commented that patients' expectation, users' expectations and satisfaction are their motivations. One respondent mentioned that developing secure mHealth apps also proves organization's proficiency to deliver security-enabled software. **R4** claimed that, *"I develop secure applications, and share our experience, so we can grow a stronger development community within our local market"*. In fact, 81% of the respondents agreed that maintaining the reputation of their organization is a motivation to developing secure mHealth apps. While 17% disagreed, only 2% were neutral to the statement SM2 in Fig. 5.

**SM3) Insecure mHealth apps have Consequences:** Monitoring patients' health, sending data to health providers, and receiving health professional decisions is one of the central features of mHealth apps. The consequence of tampered data by unauthorized entities can be damaging. Considering the consequences of insecure mHealth apps on patients's personal and health critical information received the highest mutual agreement among the respondents. 85% of the respondents (i.e., 82 out of 97 with 50% as strongly agreed and 35% as agreed) affirmed that mHealth apps suffer from negative comments and user dissatisfaction if they are insecure. Only 12% disagreed and 3% remained neutral as SM3 in Fig. 5. Respondents emphasised that safety and privacy of patients' data is their motivation to ensure the security of mHealth apps. Specifically, **R93** claimed that *"[...] and ensuring safety is our principle and bottom line"* and **R97** claimed that, *"[...], I care more about their safety"*. Two respondents **R73**, and **R77** indicated that avoiding health data leakage is their motivation to security.

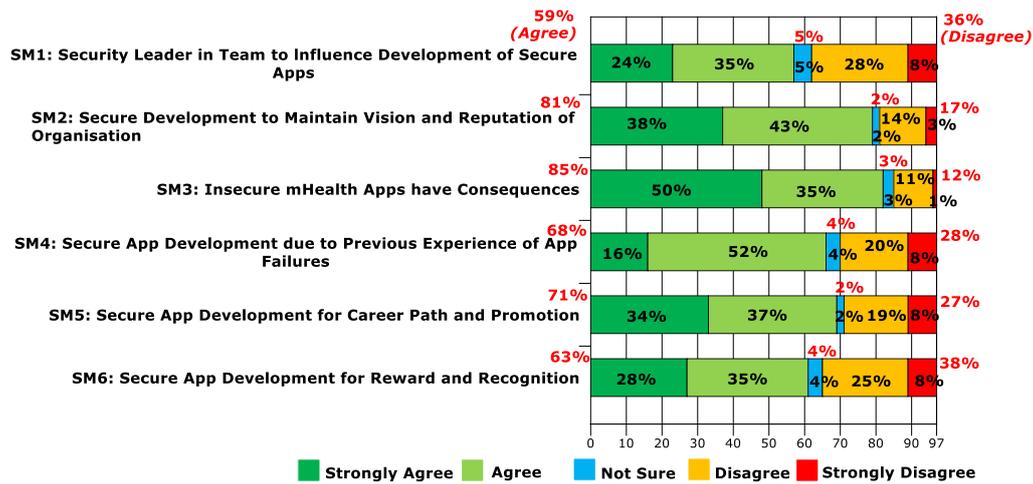

Fig. 5. Responses for Developers' Motivations to Develop Secure mHealth Apps.

**SM4) Secure App Development due to Previous Experience of App Failure:** We also wanted to examine the motivation for secure app development due to past experiences of app failure(s). 68% of the respondents agreed that they had experienced application failure in the past. Thus, it became their motivation to ensure security for the apps that they develop and specifically for mHealth apps. 28% disagreed with this statement and 4% of the respondents were neutral as SM4 in Fig. 5.

**SM5) Secure App Development for Career Path and Prmotion**: Despite the overwhelming challenges of developing secure mHealth apps, some developers puruse the development of secure mHealth apps as an ambitious activity for the sake of their intesrets and ambition. Three respondents, **R30**, **R47**, and **R63**, pointed out that personal interest, such as continuous learning and promotion is their motivations to ensure security. For example, **R47** stated that "*[I] Keep learning to improve skills [for security aware app development]*". 71% of the respondents affirmed that they develop secure mHealth apps because they look for career path and promotion with skills and expertise in software security in and secure app development in particular. 27% disagreed and 2% were neutral about SM5 in Fig. 5.

**SM6: Secure App Development for Reward and Recognition**: It is considered by developers as a perconal achievements in terms of financial or other gains that help them advance their development portfolio and profile (relevant to SM5). Expecting a reward and recognition was a motivation for developing secure mHealth apps for 63% of the respondents who agreed with SM6 in Fig. 5. 33% disagreed and 4% were neutral that reward and recognitaion is a motivation for them to do secure app development.

**Other Motivations (Open-ened):** Respondents were asked about 'other' motivating factors they see as critical but were not presented in the survey questions.

*a) Organizational practices for ensuring security:* Developing secure mHealth apps cannot be achieved without organizational commitment such as providing sufficient security tools, engaging mHealth apps developers in security training, and employing security team leads [17]). Such a commitment helps to overcome the challenges, also mentioned in Section 4A. More importantly, it would strengthen the security culture of mHealth apps developers' and their motivations to ensure security. Our respondents provided us with several responses indicating that their organizations are committed to security. **R38** indicated that proper quality assurance testing is a motivation to develop secure mHealth apps. Respondents **R31** and **R33** mentioned that they were motivated by utilizing security tools which have been provided to them by their organization.

*b) Ethical obligations:* Software development is an intellectual and effort intensive process which can be influenced by developers' behaviour. Also, mHealth apps developers' behaviour can affect their security practices (e.g., making a decision to use third-party service to process users' data), and thus, impacting the security of an app. Thus, incorporating ethical perspective is a key to motivate mHealth apps developers' to ensure security. 11 respondents **R19**, **R26**, **R37**, **R43**, **R45**, **R55**, **R59**, **R68**, **R88**, **R95**, **R96** believed that ensuring security is part of their ethical obligations. **R37**, **R43**, and **R96** indicated that they were responsible for the security of their apps. **R37** claimed that, "*Just to help everyone if I do not do my job, why would I get paid?*", and **R45** suggested, "*The main factor which motivates me to develop secure mobile health apps is that, we need to keep in mind the security of the data which a user allows to share with the application and it's duty of the development organization to keep that data private and secure that no one can misuse the data.*".

*c) Legal obligations:* Developers of mHealth apps should take into consideration the safety and privacy of their users as well as complying with laws and regulations. **R3**, **R39**, **R48** indicated that legal liability or its consequences are primary motivations to ensure security. Underestimating security can lead to breach of policies and regulations resulting in governmental fines for data breaches. Failure to secure patients' data could result in legal liability against the development organizations. **R3** indicated, "*I develop a secure mobile application because I know if any PHI [Personal Health Information] has been leaked, my organization has to pay millions of dollars*".

*d) Reducing the cost of maintaining the app:* Fixing security errors after developing mHealth apps can be error-prone, costly, and time consuming [31]. Reducing the cost of app maintenance was perceived as a motivation to ensure the security of mHealth apps, responded by **R40**.

## VII. DISCUSSION OF RESULTS AND FUTURE WORK

We now discuss the key results and highlight possible future research that extends findings of this study.

*A. Challenges for Secure mHealth App Development (RQ1)*

– Budget for Security in SDLC: Developers' perspective (i.e., 85% of respondents as in Fig. 3) highlighted that allocating little or no budget to support security specific activities in SDLC is major concern for mHealth apps.

– Insufficient SK: Knowledge of security mechanisms and implementations in development teams or at organization level hinders the development of secure mHealth apps, responded by 80% of the developers. Management support for the developers during apps development is extremely important to ensure functional as well as quality specific requirements that are related to security of mHealth apps.

**Needs for future research:** Our analysis of the security challenges for mHealth apps pinpointed the need to go beyond developers' view to also collect and analyze users' knowledge, perception, and interests towards security and privacy of their data. As part of the future work, we aim to conduct users' survey that can unveil security challenges from their perspectives (i.e., users of mHealth apps). Such a study can be beneficial to understand security from usability or users' perspective and their involvement in SDLC (Fig. 4) for secure app development.

*B. Development Practices for Secure mHealth App (RQ2)*

Based on analysis of the developers' responses, we have created a taxonomy of the adopted or recommended practices for different tasks of SDLC that represent guidelines for secure mHealth app development (Fig. 4).

– Lack of Interest and Awareness in Security by Users: 62% of our respondents claimed that users are not explicitly interested or aware of security; hence, the developers' compromise security related requirements in favour of usability and performance. Also, most of the mHealth apps stakeholders such as health professionals are non-expert in security to understand the consequences of non-secure apps. Their involvement in SDLC tasks such as requirements engineering and design for secure app development can be seen as a practice that enhances app security and stakeholders' knowledge about security issues.

**Needs for future research:** Our taxonomy of the developers' recommended practices in Fig. 4 suggests further investigation on the impact of the identified practices on secure mHealth apps. There is also a need to educate developers and disseminating knowledge about best practices and patterns to improve development of secure apps.

*C. Motivating Factors to Develop Secure mHealth App*

– Consequences of Insecure mHealth Apps: Developers' perspective (i.e., 85% respondents in Fig. 5) suggested that top factor that motivates secure mHealth app development is to avoid any legal, social, and commercial consequences of deploying insecure apps.

– Vision and Reputation of Organization: Developers are also motivated to deliver secure mHealth apps to maintain the vision and reputation of the development organization they work for, suggested by 81% of the respondents.

– Security Leads to Influence Secure SDLC: Dedicated team leads or security experts influence app developers to ensure security throughout SDLC. These leaders inspire and guide developers to satisfy security requirement and avoid security risks. However, 36% of the respondents disagreed and suggested that security leaders had not motivated or affected their development process and practices. To further analyze the rationale for their suggestion, we looked into their demographic information (i.e., team size from Fig. 2) and [26] to know that 50% of those respondents work as part-time of a team that consists of eight team members at maximum. This indicates that possibly there is a lack of security leadership within the app development team.

**Needs for future research:** There is a need for future work to investigate ethical considerations (as a motivating factor) for mHealth apps developer, especially when dealing with stakeholders who are not explicitly interested in security. Such an examination would explore the role of developers with social and legal responsibilities of secure apps development. Further work can be done to explore the role of security leaders for ensuring the security, and how they affect and motivate team members.

## VIII. THREAT TO VALIDITY

**Threat I – Source and Analysis of Survey Data:** We acknowledge that using one source for data collection (i.e., online survey) may affect our results. We believe that we did not achieve data triangulation that potentially reduces accuracy of the findings. Another possible threat relates to sampling of respondents. We recruited the developers who have experience of developing mHealth apps (Fig. 2, Section III). We explained the eligibility characteristics in survey preamble and all other recruitment documents (e.g., email invitation, post description). We asked the respondents to provide us with a concrete example of mHealth apps which they have developed to ensure their experience and eligibility. Furthermore, since we offered Amazon gift cards for respondents who provided us with eligible and complete response, there can be a threat of multiple responses by same respondent, especially those who have developed more than one mHealth app. Our study may be affected as qualitative data collection and analysis may cause irrelevant themes and misinterpretation. To overcome this threat, the first author performed the analysis and created initial codes. The third author assessed the generated codes, followed by a discussion to confirm the final themes (Section III).

**Threat II – Survey Data Collection and Sample Size:** Questionnaire-based research faces the risk of being misunderstood or misinterpreted by respondents. We conducted pilot survey with questionnaire to overcome wording and ambiguity issues (as in Fig. 2). We made some questions compulsory to answer. It was hard to reach respondents who responded earlier to seek their inputs for the compulsory questions. After sending several emails, we only received a few replies. However, we believe that our study has a convenience sample (i.e., N=97) based on available pool of mHealth apps developers. All our respondents have at least developed one mHealth app, are geographically diversified, and have different expertise and team size that reflects minimum knowledge and expertise to respond.

## IX. CONCLUSIONS

Mobile computing empowers health care stakeholders – offering portable, context-sensitive, and pervasive computing – to produce and consume healthcare services [1]. Mobile apps in general and mHealth apps in particular face critical challenges related to security and privacy of users' information and health critical data. We conducted an empirical study by collecting, analyzing, synthesizing, and documenting responses about secure app development from 97 mHealth app developers from across the World. The study is primarily focused on three objectives expressed as RQs, i.e., (i) *what are the challenges*? (ii) *which development practices are critical*? and (iii) *how motivating factors impact* the development of secure mHealth apps. The results of this study can benefit researchers and practitioners with empirical knowledge about security specific challenges and opportunities to develop secure mHealth apps.